\documentclass[12pt]{iopart}

\usepackage{graphicx}
\usepackage{aas_macros}
\usepackage{hyperref}
\usepackage{bm}
\usepackage{color}

\usepackage{caption}
\usepackage{subcaption}

\begin{document}

\title[Efficient Bayesian inference and model selection for CWs in PTA data]{Efficient Bayesian inference and model selection for continuous gravitational waves in pulsar timing array data}

\author{Bence B\'ecsy}

\address{Department of Physics, Oregon State University, Corvallis, OR 97331, USA}
\ead{becsyb@oregonstate.edu}
\vspace{10pt}

%\author{TBD}

%\address{places}
%\begin{indented}
%\item[]August 2017
%\end{indented}

\begin{abstract}
Finding and characterizing gravitational waves from individual supermassive black hole binaries is a central goal of pulsar timing array experiments, which will require analysis methods that can be efficient on our rapidly growing datasets. Here we present a novel approach built on three key elements: i) precalculating and interpolating expensive matrix operations; ii) semi-analytically marginalizing over the gravitational-wave phase at the pulsars; iii) numerically marginalizing over the pulsar distance uncertainties. With these improvements the recent NANOGrav 15yr dataset can be analyzed in minutes after an $\mathcal{O}(1\ \mathrm{hour})$ setup phase, instead of an analysis taking days-weeks with previous methods. The same setup can be used to efficiently analyze the dataset under any sinusoidal deterministic model. In particular, this will aid testing the binary hypothesis by allowing for efficient analysis of competing models (e.g.~incoherent, monopolar, or dipolar sine wave model) and scrambled datasets for false alarm studies. The same setup can be updated in minutes for new realizations of the data, which enables large simulation studies.
\end{abstract}

%\maketitle

\section{Introduction}
\label{sec:intro}

With recent pulsar timing array (PTA) datasets \cite{nanograv_15yr_data, epta_dr2_data, ppta_dr3_data} showing evidence for a stochastic gravitational-wave (GW) background (GWB) \cite{nanograv_15yr_gwb, epta_dr2_gwb, ppta_dr3_gwb, cpta_dr1_gwb}, the nHz GW sky promises to be an exciting discovery space in the next decades. An important next milestone will be determining the source of the GWB. The most obvious candidate is a superposition of signals from supermassive black hole binaries (SMBHBs) \cite{nanograv_15yr_astro, epta_astro}, but more exotic sources cannot be ruled out at this point \cite{nanograv_15yr_new_physics, epta_astro}. 

If the GWB is coming from SMBHBs, we can expect to individually resolve the loudest of those binaries within the next decade \cite{SVV2009, RosadoExpectedProperties, ChiaraLandscape,Luke_single_source,RealisticDetection1}. This will not only help us unequivocally pinpoint the source of the GWB, but these sources would be of great interest in their own right. They are expected to be rich multi-messenger sources, and they could help us better understand the formation and evolution of massive black holes \cite{Bogdanovic_SMBH_EM,Kelley_white_paper_2019,Charisi+2022}. To realize these sources' potential we need efficient and robust analysis methods to search for their signals, characterize them, and distinguish them from other potential noise sources and signals. These are challenging tasks due to the large data volumes, and unevenly sampled datasets that necessitate a time-domain analysis.

Several previous searches have been carried out to look for these sources, setting increasingly stringent upper limits over the years \cite{yardley+10, nanograv_5yr_cw, PPTA-cw-paper, EPTA-CW-paper, nanograv_11yr_cw, nanograv_12p5yr_cw, ipta_dr2_cw, nanograv_15yr_cw, epta_dr2_cw}. Most recently, Ref.~\cite{nanograv_15yr_cw} and Ref.~\cite{epta_dr2_cw} searched for individual binaries in the NANOGrav 15-year \cite{nanograv_15yr_data} and EPTA DR2 \cite{epta_dr2_data} datasets, respectively. While these found no strong evidence for a resolvable binary, they uncovered interesting candidates. This highlights the importance of rigorous model selection and additional checks if we want to reliably vet such candidates in the future.

As our datasets grow, and the presence of a stochastic background becomes more clear, analyses have to be improved to be efficient and allow for searching for an individual binary in the presence of a background. Numerous Bayesian analysis techniques have been developed to search for and characterize GWs from individual SMBHBs (see e.g.~\cite{NeilCWMethods, Lee_et_al_CW_methods, JustinCWMethods, Steve_accelerated_CW, QuickCW}). Most recently, the \texttt{QuickCW} pipeline \cite{QuickCW} introduced a new formulation of the likelihood that allows more expensive calculations to be done less frequently, thus providing an overall speedup of about a factor of 100. While this is sufficient to carry out a search in a few days (instead of months) on the most recent PTA datasets, it is still not fast enough to allow for other analyses requiring more complicated setups or many independent runs, e.g.: i) simulation studies on thousands of realizations; ii) false alarm estimation of candidates via scrambled datasets similar to techniques used for the GWB (see e.g.~Refs.~\cite{neil_robust_detection,all_correlations_must_die}); iii) comparing an individual binary model with alternative sinusoidal models (e.g.~incoherent, monopolar, or dipolar sine models); iv) multiple binaries \cite{babak_sesana_multiple_cw,BayesHopper}; v) non-circular binaries \cite{steve_eccbin,abhimanyu_eccbin_2020,abhimanyu_eccbin_2022,nanograv_12p5yr_eccbin}; vi) beyond-general-relativity binaries \cite{logan_alt_pol_cw}; etc.

In this paper we present a method implemented in the \texttt{FürgeHullám} package\footnote{\url{https://github.com/bencebecsy/FurgeHullam}}\footnote{The name comes from Hungarian words "fürge" (meaning quick) and "hullám" (meaning wave).} that is an additional factor of 10-100 times faster than Ref.~\cite{QuickCW}, as long as the noise parameters are fixed. This can be sufficient for many of the use cases, and can potentially be relaxed in a future implementation. However, this means that \texttt{QuickCW} can be thought of as a full-fledged search pipeline, while \texttt{FürgeHullám} can either be used as a parameter estimation and model selection tool or as a first-pass quick search. This speedup is achieved via the combination of three different techniques:
\begin{enumerate}
    \item Precalculating and interpolating expensive matrix operations (see Section \ref{sec:interpolation});
    \item Semi-analytically marginalizing over the gravitational-wave phase at the pulsars (see Section \ref{sec:phase-marg});
    \item numerically marginalizing over the pulsar distance uncertainties (see Section \ref{sec:distance-marg}).
\end{enumerate}
Note that (ii) and (iii) are similar to techniques presented in Ref.~\cite{Steve_accelerated_CW}. For their main results they use numerical phase marginalization, which our results improve upon by deriving analytic expressions, which are much faster. In addition, they also derive an analytic phase marginalization similar to ours. However, their derivation makes several assumptions that we relax in this paper, for example that there is no frequency evolution and that the filter functions are orthogonal. We relax these assumptions, which promotes the analytic marginalization technique from a quick but biased analysis to an efficient and accurate one. Ref.~\cite{iterative_multiCW_yiqian2022} also uses pulsar phase marginalization, however, they also carry out the marginalization numerically, albeit in a more sophisticated way that ensures numerical stability at various signal strengths.

The rest of the paper is organized as follows. In Section \ref{sec:methods} we review the signal model and describe the methods used in this new analysis. In Section \ref{sec:test} we validate this approach on simulated datasets. Finally, we offer concluding remarks and discuss future directions enabled by this work in Section \ref{sec:conclusion}.

%\redtext{TODO: figure out when "no frequency evolution between psr and earth term" and "no frequency evolution within observing timespan" approximations are okay. Use match between true waveform and these over various frequencies and  plot regions where these are okay. In addition, we can also look at holodeck population to see the fraction of detectable binaries that fall under these regions. Instead just quote the match for the example source in this paper!}

%\redtext{TODO: maybe try nested sampling where some parameters are sampled with slice sampling, while others (like psr distance and maybe phase) are simply drawn from the prior}

\section{Methods}
\label{sec:methods}

It has been recognized in previous studies, that the GW signal of a circular SMBHB in PTA data can be decomposed into a sum over four filter functions: a sine and cosine describing the GW signal at the observer's location (\emph{Earth term}) and a sine and cosine describing the signal at the pulsar location (\emph{Pulsar term}). Here we give a brief overview of the signal model in this formulation (for more details see Ref.~\cite{QuickCW}). The signal contributing to the times-of-arrival (TOAs) of the $\alpha$th pulsar is:
\begin{equation}
    s_{\alpha} = \sum_{i=1}^4 b_{i\alpha} S^i_{\alpha},
    \label{eq:signal_decomp}
\end{equation}
where $S^i_{\alpha}=[\sin (2 \pi f_{\rm E} t); \cos (2 \pi f_{\rm E} t); \sin (2 \pi f_{\alpha} t); \cos (2 \pi f_{\alpha} t)]$ are the four necessary filter functions at the Earth-term ($f_{\rm E}$) and pulsar-term ($f_{\alpha}$) frequencies. Here we make the realistic assumption that there is no frequency evolution during the $\mathcal{O}(10 \ {\rm year})$ observing window, but there can be significant frequency evolution during the $\mathcal{O}(10^3 \ {\rm year})$ light-travel time between the Earth and the pulsar. This is a good assumption for most realistic scenarios (see Section \ref{sec:test} for more justification), but can also be relaxed in the future\footnote{This can be done e.g.~by introducing an extra parameter in these filters and thus increase the dimensionality of the interpolation discussed in Section \ref{sec:interpolation}; or by decomposing the chirping signal as a sum of multiple sine waves, thus increasing the number of filters in Eq.~(\ref{eq:signal_decomp}).}. The reason we need both a sine and a cosine is to be able to combine these to get any possible phase shift. The $b_{i\alpha}$ coefficients can be written as:
\numparts
\begin{eqnarray}
    b_{1\alpha} &= A \, , \\
    b_{2\alpha} &= B \, , \\
    b_{3\alpha} &= \left( \frac{f_{\rm E}}{f_{\alpha}} \right)^{1/3} \left[ -A \cos \Phi_{\alpha} -B \sin \Phi_{\alpha} \right] \, , \\
    b_{4\alpha} &= \left( \frac{f_{\rm E}}{f_{\alpha}} \right)^{1/3} \left[ -B \cos \Phi_{\alpha} +A \sin \Phi_{\alpha} \right] \, ,
    \label{eq:b_coeffs}
\end{eqnarray}
\endnumparts
where $\Phi_{\alpha}$ are is the phase of the pulsar at the $\alpha$th pulsar's location, and:
\numparts
\begin{eqnarray}
    A &= F^{+}_{\alpha} a_1 + F^{\times}_{\alpha} a_3\, , \\
    B &= F^{+}_{\alpha} a_2 + F^{\times}_{\alpha} a_4 \, ,
\end{eqnarray}
\endnumparts
where $F^{+/\times}_{\alpha}$ are the antenna patterns for the $\alpha$th pulsar (defined e.g.~in Eq.~(7) and (8) in Ref.~\cite{QuickCW}), and $a_i$ are coefficients defined as:
\numparts
\begin{eqnarray}
    a_1 &= -A_{\rm e} (2 \pi f_{\rm E})^{-1} \left[ \cos 2\Phi_0 (1+\cos^2 \iota) \cos 2\psi - 2 \sin 2\Phi_0 \cos\iota \sin 2\psi\right] \, ,   \\
    a_2 &= -A_{\rm e} (2 \pi f_{\rm E})^{-1} \left[ \sin 2\Phi_0 (1+\cos^2 \iota) \cos 2\psi + 2 \cos 2\Phi_0 \cos\iota \sin 2\psi \right] \, ,   \\
    a_3 &= A_{\rm e} (2 \pi f_{\rm E})^{-1} \left[ \cos 2\Phi_0 (1+\cos^2 \iota) \sin 2\psi + 2 \sin 2\Phi_0 \cos\iota \cos 2\psi \right]\, ,   \\
    a_4 &= A_{\rm e} (2 \pi f_{\rm E})^{-1} \left[ \sin 2\Phi_0 (1+\cos^2 \iota) \sin 2\psi - 2 \cos 2\Phi_0 \cos\iota \cos 2\psi \right] \, ,
    \label{eq:a_coeffs}
\end{eqnarray}
\endnumparts
where $A_{\rm e}={\cal M}^{5/3} d_L^{-1} (2 \pi f_{\rm E})^{2/3}$ is the Earth-term signal amplitude, which is determined by the observer-frame chirp mass of the binary ($\mathcal{M}$), the luminosity distance to the source ($d_L$) and the observer-frame Earth term GW frequency ($f_{\rm E}$). The $a_i$ coefficients are also influenced by the initial phase of the Earth-term signal ($\Phi_0$), the inclination angle of the binary's orbit ($\iota$) and the GW polarization angle ($\Psi$). In addition, $F^{+/\times}_{\alpha}$ depends on the sky location of the binary ($\theta$ and $\Phi$), and the pulsar terms are influenced by the GW phase ($\Phi_{\alpha}$) and observer-frame GW frequency ($f_{\alpha}$) at the pulsar's location. In principle $\Phi_{\alpha}$ and $f_{\alpha}$ are determined by the distance to the given pulsar ($L_{\alpha}$). However, in practice $L_{\alpha}$ are not known precisely enough, so we include them as parameters with priors informed by our current distance measurements (see Section \ref{sec:distance-marg} for more details). These together result in a total of $8+2N_{\rm PSR}$ parameters that describe the signal, where $N_{\rm PSR}$ is the number of pulsars in the array.

This decomposition can be used to form a maximum-likelihood estimator, called the F-statistic \cite{PTAFstat}. A similar decomposition is also used in the \texttt{QuickCW} pipeline to separate parameters that change the $S^i_{\alpha}$ filter functions (expensive to update shape parameters) and ones only changing the $b_{i\alpha}$ coefficients (cheap to update projection parameters) \cite{QuickCW}. Note that the decomposition described above is slightly different from the one used in \texttt{QuickCW} \cite{QuickCW}. In this work we do not include frequency evolution within the observing timespan, while \texttt{QuickCW} does take that into account.

\subsection{Inner product spline interpolation}
\label{sec:interpolation}

Central to any Bayesian analysis is the log likelihood function, which in this case can be written as:
\begin{equation}
    \ln L = \sum_{\alpha=1}^{N_{\rm PSR}} (d_{\alpha}-s_{\alpha}|d_{\alpha}-s_{\alpha}) = \sum_{\alpha=1}^{N_{\rm PSR}} \left[ \sum_{i=1}^4 b_{i\alpha} N^i_{\alpha} - \frac{1}{2} \sum_{i=1}^4 \sum_{j=1}^4 b_{i\alpha} b_{j\alpha} M^{ij}_{\alpha} \right],
    \label{eq:log_like}
\end{equation}
where $d_{\alpha}$ is the data from the $\alpha$th pulsar, $N^i_{\alpha} = (d_{\alpha}|S^i_{\alpha})$, and $M^{ij}_{\alpha}=(S^i_{\alpha}|S^j_{\alpha})$, and $(a|b)=a^TC^{-1}b$ denotes the inner product between two vectors with some $C$ covariance matrix that includes all the information about the noise model for the given pulsar. Note that as long as there are no noise processes correlated between pulsars, the log likelihood factorizes as a sum over terms for different pulsars. However, the GWB, which started to show up in recent PTA datasets, is such a correlated process, so it will be important in the future to include inter-pulsar correlations. Potential ways to extend this work to include correlated noise is discussed in Section \ref{sec:conclusion}. We can also see that the log likelihood is a sum of various inner products between the data and the filter functions ($N^i_{\alpha}$) and between filter functions ($M^{ij}_{\alpha}$). By far the most expensive operation in calculating the likelihood is evaluating these inner products, after which we get the likelihood with just a couple of additions and multiplications. \texttt{QuickCW} makes use of this fact by only occasionally updating the parameters of the signal model that affect the filter functions (half of the parameters). This allows extremely quick exploration of parameters that only affect the coefficients, and thus speeds up the entire search.

\begin{figure}[!htbp]
 \centering
    \includegraphics[width=0.65\textwidth]{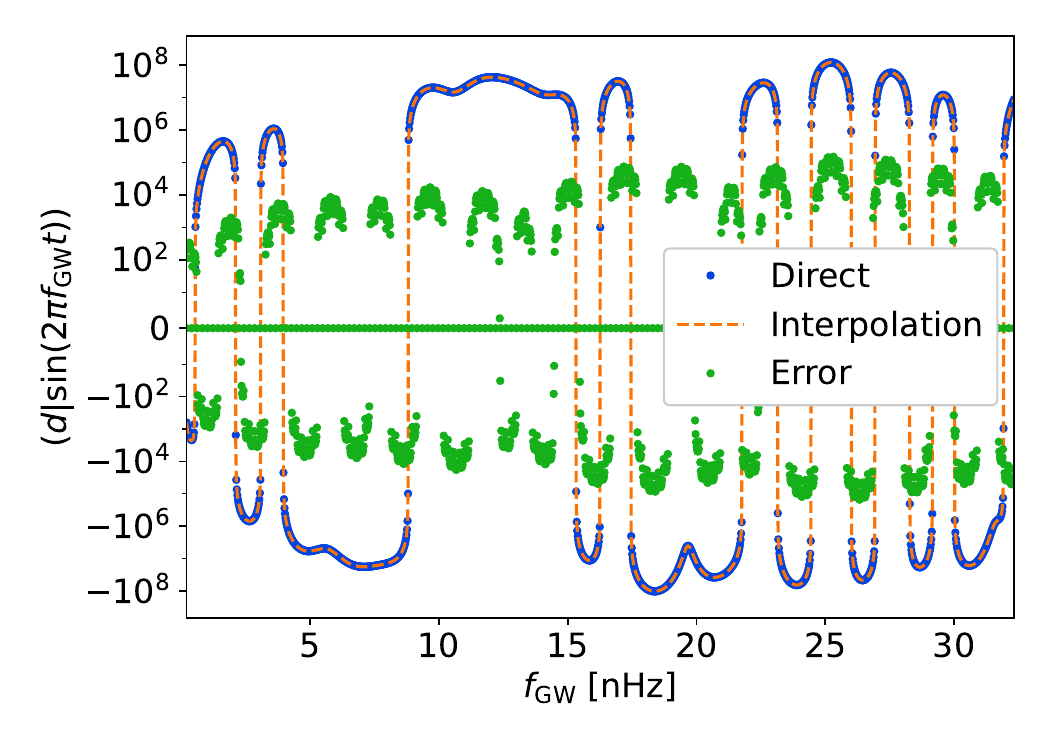}
    %\fakeimage
 \caption{Inner products of a sine with various frequencies with the data (blue dots) for a simulated dataset made to resemble the data of PSR B1855+09 from the NANOGrav 15yr dataset. Also shown are the local cubic interpolant (orange dashed curve) and the absolute error of the interpolation (green dots).}
 \label{fig:N_interpolation}
\end{figure}

Here we introduce a different way of utilizing this structure of the likelihood, by precalculating these inner products on a grid and using interpolation to get their value at any location in parameter space. This approach was used in Ref.~\cite{BWM_lookup_table} for analyzing a signal from a burst with memory. However, the binary signal model has more parameters, thus a grid in those parameters would be prohibitively expensive. However, we can notice that while half the signal parameters affect $N^i_{\alpha}$ and $M^{ij}_{\alpha}$, they only do so by changing the frequency of these filter functions. Thus fundamentally, $N^i_{\alpha}$ and $M^{ij}_{\alpha}$ are only functions of the Earth term and pulsar term frequencies, which are in turn functions of other model parameters, like the pulsar distances, chirp mass, and sky location. This means that we can precalculate these inner products on a grid and interpolate between them whenever we need $N^i_{\alpha}$ and $M^{ij}_{\alpha}$ at particular frequencies. While $N^i_{\alpha}$ has 4 components for each pulsar, the Earth term and pulsar term filters are the same except they are at different frequencies. So in practice we only need two interpolations, one for a sine and one for a cosine with various frequencies. Figure \ref{fig:N_interpolation} shows the inner products of sine with the data as a function of frequency (blue dots) for simulated data made to resemble the data of PSR B1855+09 from the NANOGrav 15yr dataset \cite{nanograv_15yr_data}\footnote{The corresponding plot for the inner product between cosine and data looks qualitatively the same.}. Note that while this may not be immediately apparent due to the symmetric log scale used for this plot, this function is fairly smooth, which allows for simple low-order interpolation. We also show a local cubic interpolation (orange dashed curve), which is based on a grid evenly sampled in frequency with a spacing of $1/(10T_{\rm obs})$, where $T_{\rm obs}$ is the total observing timespan. We found this spacing to be the least number of points needed for reliable results. The absolute error of the interpolation (green points) is also shown in Figure \ref{fig:N_interpolation}. All of these were evaluated on a grid 10 times finer than the one used for the interpolation. Note that the errors are typically orders of magnitudes smaller than the actual values\footnote{The green points along zero correspond to points where the interpolant was evaluated at the same point where it was precalculated, so the error is at much smaller scales and corresponds to numerical precision.}.

\begin{figure}[!htbp]
 \centering
    \begin{subfigure}[b]{0.5\textwidth}
            \centering
            \includegraphics[width=1.05\linewidth]{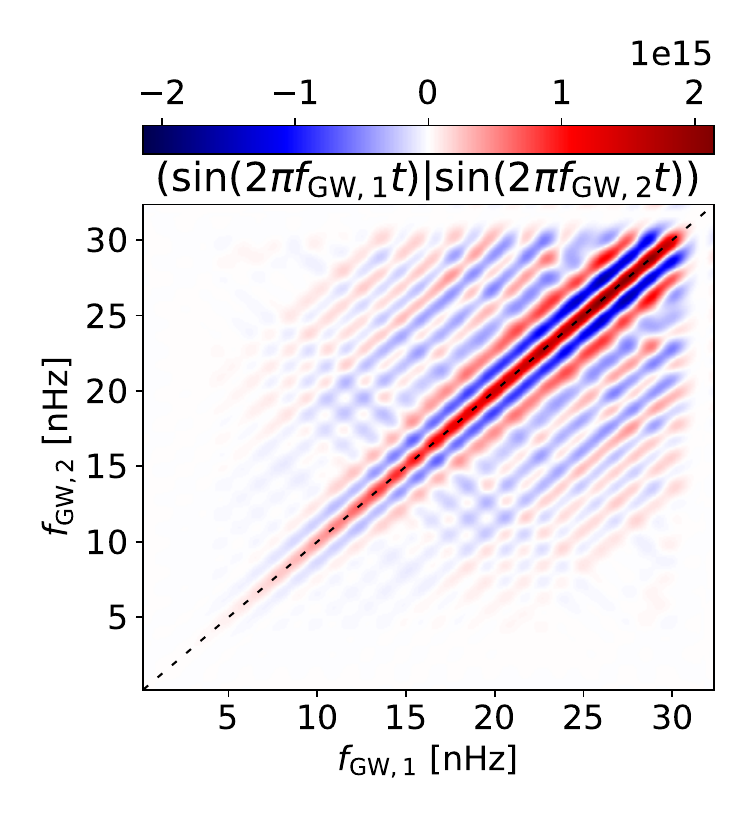}
            %\fakeimage
            \caption{}
    \end{subfigure}%
    \hfill
    \begin{subfigure}[b]{0.5\textwidth}
            \centering
            \includegraphics[width=1.05\linewidth]{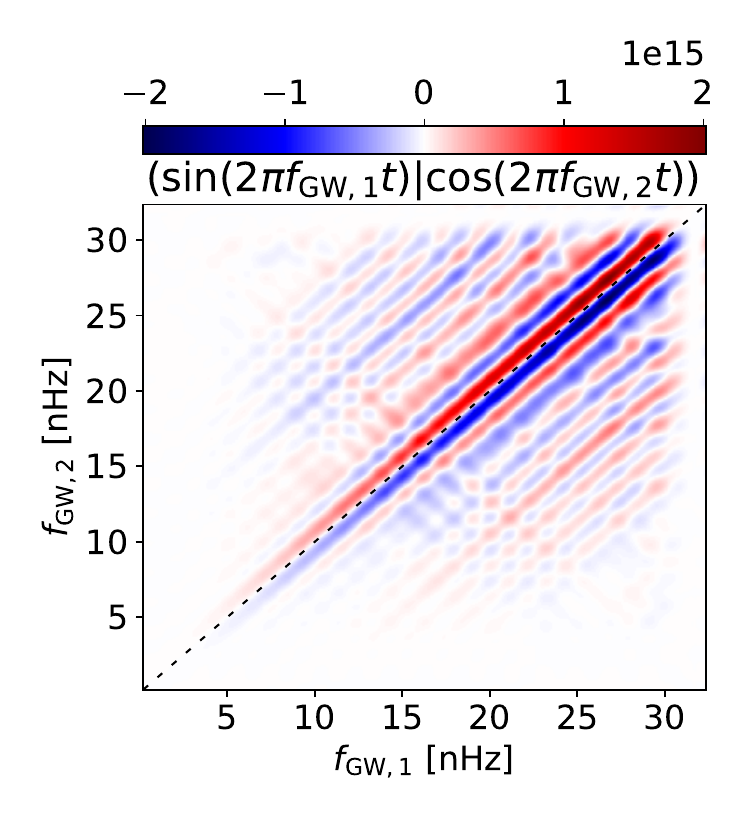}
            %\fakeimage
            \caption{}
    \end{subfigure}%
    \hfill
 \caption{Inner products between sine and sine (a) and sine and cosine (b) as a function of frequency for a simulated dataset made to resemble the data of PSR B1855+09 from the NANOGrav 15yr dataset. Dashed lines show $f_{{\rm GW},1}=f_{{\rm GW},2}$.}
 \label{fig:M_interpolation}
\end{figure}

Now consider $M^{ij}_{\alpha}$, which is a 4-by-4 symmetric matrix for each pulsar, thus having 10 independent components. However, all of these fall into 3 different categories: i) sine-sine; ii) cosine-cosine; iii) sine-cosine. Thus we only need three distinct 2-dimensional interpolants. These are constructed similarly to those for $N^i_{\alpha}$, but on a 2-dimensional grid, using a 2-dimensional local cubic interpolation. Figure \ref{fig:M_interpolation} shows the sine-sine and sine-cosine interpolants\footnote{The cosine-cosine interpolant looks qualitatively similar to the sine-sine interpolant shown.}. Note that as expected, the sine-sine term is maximal when the frequencies are the same, and shows a sinc-function-like decay for different frequencies. Similarly, the sin-cos term is close to zero for equal frequencies and shows a periodic decaying pattern as the frequency difference increases. However, the uneven sampling and non-stationary noise results in non-trivial modification of these simple expectations. For example, frequencies close to zero and $1/{\rm yr}\simeq32$ nHz show inner products close to zero, as the timing model fit down-weights frequencies covariant with the timing model. There is also an overall upward trend as frequency increases, due to red noise that reduces the noise-weighted inner products at low frequencies. Also note that the sine-cosine inner products are comparable in magnitude to the sine-sine inner products, and depending on the frequencies they can even be larger. This highlights that these cross-terms cannot be ignored in this case, even though they would be zero if the data were evenly sampled and we were only looking at multiples of the fundamental $1/T_{\rm obs}$ frequency. Similar to the $N^i_{\alpha}$ interpolants, $M^{ij}_{\alpha}$ interpolants typically show relative errors of less than 1\%.

Thus we can see that in total we need two 1-dimensional, and three 2-dimensional interpolants. We found the reliable results can be achieved with 10 grid points per $1/T_{\rm obs}$-wide frequency bin. So if we want to analyze up to a frequency of $N/T_{\rm obs}$, we need $10N$ inner product calculations for each 1-dimensional grids, $100N^2$ inner products for the sine-cosine inner products, and $5N(1+10N)$ for the sine-sine and cosine-cosine grids each, because those are symmetric. Thus in total, we need $30N+200N^2$ inner product calculations. For the realistic scenario where we want to search up to a frequency of 1/yr in a 16 yr long dataset ($N=16$), the interpolation setup requires $\sim$52 thousand inner product calculations. Note however, that one can save on the 2-dimensional grids by first calculating and storing $C^{-1}b$ for all $10N$ $b$ vectors and then contracting that with the $10N$ $a^T$ vectors. Thus in practice the setup is about as costly as 20 thousand likelihood calls, which is much faster than a naive Markov-chain Monte Carlo (MCMC) that calculates the likelihood from scratch each time. We can see in Table \ref{tab:timings} that this setup on a simulated dataset with the same data volume as the NANOGrav 15yr dataset takes about 4.5 hours on a modest 16-core CPU. In addition, unlike a serial MCMC, this setup is trivially parallelizable, since we already know each inner product we wish to calculate. Thus a setup time of $<$1 hour is easily possible, but utilizing the high parallel-computing capabilities of GPUs could reduce the setup time to a few minutes (see e.g.~\cite{DiscoveryGPU} for a recent effort to parallelize PTA computations on GPUs). Also note that the majority of the setup is spent on the 2-dimensional interpolants, which do not depend on the data. Thus if we want to analyze many realizations of simulated datasets, where the data changes, but the covariance matrix and observing epochs stay the same, the setup only costs $20N$ inner products, or about 3.5 minutes instead of 4.5 hours in the above example.

\begin{table*}[htbp]%The best place to locate the table environment is directly after its first reference in text
\centering
\caption{%
Representative evaluation times of different variants of the likelihood on a NANOGrav 15yr-like dataset (67 pulsars, 674,683 TOAs) with an AMD Ryzen Threadripper 3970X 32-core processor (restricted to use only 16 cores). Note that all three approaches use the same model, which includes white noise, pulsar red noise, an uncorrelated common red noise, analytical marginalization over the timing model uncertainties, and an evolving individual binary with pulsar terms included. \label{tab:timings}
}
\begin{tabular}{|l|l|}
\hline
Enterprise likelihood & 1.4 s\\
\hline
QuickCW fast step & 15 $\mu$s\\ %18.6 us (from 100 steps) - 11.75 us (from 1k steps) - 11 us
QuickCW slow step & 500 ms\\ %0.75 s (from 100 steps) - 0.53 s for intrinsic, 0.93 s for distance - 0.57 s (from 1k run) - 0.54 s
%100k slow-step QuickCW run & 50 h\\
\hline
Unparallelized setup time & 4.5 h\\
Setup update with new data & 3.5 min\\
Splined likelihood & 350 $\mu$s \\
Phase-marginalized splined likelihood & 500 $\mu$s\\
Phase-and-distance-marginalized splined likelihood & 3.5 ms \\
\hline
\end{tabular}
\end{table*}

We tested the accuracy of the log likelihood calculated with these interpolated likelihoods on a simulated dataset made to resemble the NANOGrav 15yr dataset. We injected an individual binary signal and analyzed it with \texttt{QuickCW}. Then we randomly drew samples from the posterior and calculated the log likelihood at these points with the interpolated inner products. The absolute difference between the directly computed and interpolated likelihoods was always less than 3\%, and most of the time less than 1\%. These are larger than the expected numerical errors ($\sim10^{-5}$), but they are small enough not to affect results appreciably (see Section \ref{sec:test} for more consistency tests). Table \ref{tab:timings} also shows the average time it takes to calculate this interpolated likelihood. It is interesting to compare this evaluation time with the traditional \texttt{enterprise} \cite{enterprise} likelihood call and fast and slow \texttt{QuickCW} likelihood calls, which are also shown in Table \ref{tab:timings}. We can see that the splined likelihood is $\sim$4,000 times faster than \texttt{enterprise}, $\sim$1,500 times faster than a \texttt{QuickCW} slow step, and $\sim$25 times slower than a \texttt{QuickCW} fast step. So we can expect an analysis using this splined likelihood not only to be vastly faster than \texttt{enterprise}, but also to be significantly faster than \texttt{QuickCW}, since it can make relatively fast steps in all parameters.

%\begin{figure}[htb]
% \centering
%   \includegraphics[width=0.7\textwidth]{log_like_diff_phase_approx.pdf}
% \caption{Distribution of likelihood errors due to $N^i_{\alpha}$ and $M^{ij}_{\alpha}$ cubic interpolation on a simulated dataset made to resemble the NANOGrav 15yr dataset. While these are much larger than the expected numerical errors of direct likelihood calculation, they are small enough that they should not affect the results appreciably.}
% \label{fig:likelihood_accuracy}
%\end{figure}

The current implementation uses a local cubic interpolant based on the \texttt{fast\_interp} package\footnote{\url{https://github.com/dbstein/fast_interp}}, which utilizes just-in-time compilation with \texttt{Numba}\cite{numba_paper, numba_zenodo} to make these really fast. As we have seen this achieves reasonable accuracy and speed based on a moderately high number of interpolation points. However, it is possible that more advanced interpolation methods would result in better accuracy and speed or require fewer evaluations for a setup. One promising approach could be using Gaussian processes, which potentially could match the intrinsic periodic variability we see on Figures \ref{fig:N_interpolation} and \ref{fig:M_interpolation}. Another promising avenue would be using machine-learning-based interpolation approaches. We leave these investigations of potential further improvements to future work.

\subsection{Semi-analytic phase marginalization}
\label{sec:phase-marg}

The phases of the GW at the locations of the pulsars ($\Phi_{\alpha}$) are nuisance parameters that carry no physical information about the GW source\footnote{At least in the current situation where the error on the pulsar distances is much larger than a GW wavelength.}. Thus it could be beneficial to marginalize over them explicitly, instead of including them in the list of parameters sampled via MCMC. A similar approach was suggested for ground-based GW detectors \cite{veitch2013analytic}. PTAs could potentially see even more benefits, due to the large number of pulsar phase parameters. In addition, since a given pulsar phase parameter only affects the given pulsar, each pulsar phase can be integrated out via a 1-dimensional integral. So by explicitly integrating these parameters out, we are calculating a 1-dimensional integral $N_{PSR}$ times, instead of a single $N_{PSR}$-dimensional integral that the MCMC is doing when we leave these parameters to sample over. This suggests that even a naive numerical integral of these phase parameters is better than sampling over them with MCMC. This argument has been made in Ref.~\cite{Steve_accelerated_CW}, where they numerically marginalize over phase parameters. Ref.~\cite{iterative_multiCW_yiqian2022} also expands on this approach by a clever rescaling of the integral, which ensures numerical stability. However, as it turns out the required integral can be done analytically.

Adopting the uninformative uniform prior between 0 and $2\pi$ on the pulsar phases, the integral in question for the $\alpha$th pulsar has the following form:
\begin{eqnarray}
I_{\alpha} &= \int_0^{2\pi} \exp L(\Phi_{\alpha}) \ d\Phi_{\alpha} \nonumber\\ &= \int_0^{2\pi} \exp \left[\sum_{i=1}^4 b_{i\alpha} N^i_{\alpha} - \frac{1}{2} \sum_{i=1}^4 \sum_{j=1}^4 b_{i\alpha} b_{j\alpha} M^{ij}_{\alpha} \right]  \ d\Phi_{\alpha}.
\end{eqnarray}
Using Eq.~(\ref{eq:b_coeffs}) we can rewrite the log likelihood in the integrand to highlight its $\Phi_{\alpha}$ dependence:
\begin{equation}
    L (\Phi_{\alpha}) = L_0 + L_1 \sin (\Phi_{\alpha}) + L_2 \cos (\Phi_{\alpha}) + L_3 \sin (2\Phi_{\alpha}) +L_4 \cos (2\Phi_{\alpha}),
    \label{eq:L_of_phi}
\end{equation}
where (dropping the $\alpha$ indices):
\numparts
\begin{eqnarray}
    \fl L_0 &= A N^1 + B N^2 - \frac{A^2}{2} M^{11} - \frac{B^2}{2} M^{22} - ABM^{12} - \frac{\chi^2}{4} \left(A^2+B^2\right) \left(M^{33}+M^{44}\right) \, , \\
    \fl L_1 &= \chi \left[ -BN^3 + AN^4 + ABM^{13} - A^2 M^{14} + B^2 M^{23} -ABM^{24} \right] \, ,  \\
    \fl L_2 &= \chi \left[ -AN^3 - BN^4 + A^2 M^{13} + AB M^{14} + AB M^{23} +B^2 M^{24} \right] \, , \\
    \fl L_3 &= \frac{\chi^2}{2} \left[ AB \left( M^{44} - M^{33} \right) + \left(A^2-B^2\right) M^{34} \right] \, , \\
    \fl L_4 &= \frac{\chi^2}{4} \left[ \left(A^2-B^2\right) \left(M^{44} - M^{33}\right) - 4AB M^{34} \right] \, ,
\end{eqnarray}
\endnumparts
where $\chi=(f_{\rm E}/f_{\alpha})^{1/3}$. Note that the non-vanishing double-angle terms ($L_3$ and $L_4$) come from the fact that $M^{34}\neq0$ and $M^{33}\neq M^{44}$ due to the uneven sampling of the dataset and the fact that frequencies are not necessarily integer multiples of $1/T_{\rm obs}$. This is unlike the situation in ground-based detectors, where these terms do vanish, thus the resulting integral is simply a 0th order modified Bessel function of the first kind \cite{LIGO_CW_phasemarg,veitch2013analytic,LALInference}. Note that a phase-marginalized likelihood was also derived for PTAs under the assumption that $M^{34}=0$ and $M^{33}= M^{44}$ (see Appendix A in Ref.~\cite{Steve_accelerated_CW}). However, as we can see on Figure \ref{fig:M_interpolation}, $M^{34}$ can be comparable to $M^{11}$ and $M^{22}$, so this is not a particularly accurate approximation in realistic scenarios.

The integral then can be written as:
\begin{eqnarray}
    \fl I_{\alpha} &= e^{L_0} \int_0^{2\pi} \exp \left[ f(\Phi_{\alpha}) \right] \ d\Phi_{\alpha} \nonumber\\
%    \fl &= e^{L_0} \int_0^{2\pi} \exp \left[ \sqrt{L_1^2 + L_2^2} \cos (\Phi_{\alpha} - \tan^{-1}[L_1/L_2]) + \sqrt{L_3^2 + L_4^2} \cos(2\Phi_{\alpha} - \tan^{-1}(L_3/L_4)) \right] \ d\Phi_{\alpha}\nonumber\\
    \fl &= e^{L_0} \int_0^{2\pi} \exp \left[ C \cos (\Phi_{\alpha} - \Phi_C) + D \cos(2\Phi_{\alpha} - 2\Phi_D) \right] \ d\Phi_{\alpha}\nonumber\\
    \fl &= e^{L_0} \int_0^{2\pi} \exp \left[ x\sin (\Phi_{\alpha}') + y\cos (\Phi_{\alpha}') + z \cos(2\Phi_{\alpha}') \right] \ d\Phi_{\alpha}',
    \label{eq:integral_manipulations}
\end{eqnarray}
where we got the second line by combining the second and third, and the forth and fifth terms of Eq.~(\ref{eq:L_of_phi}) via trigonometric addition formulas resulting in the following coefficients and phases:
\numparts
\begin{eqnarray}
    C &= \sqrt{L_1^2 + L_2^2} \, , \\
    D &= \sqrt{L_3^2 + L_4^2} \, , \\
    \Phi_C &= \tan^{-1}(L_1/L_2) \, , \\
    \Phi_D &= \tan^{-1}(L_3/L_4) / 2 \, .
\end{eqnarray}
\endnumparts
Then we reach the last line of Eq.~(\ref{eq:integral_manipulations}) by shifting the integration variable to $\Phi_{\alpha}' = \Phi_{\alpha} - \Phi_D$, which gets rid of the phase offset in the double angle term and thus reduces the number of terms from four to three. The resulting coefficients can be expressed as:
\numparts
\begin{eqnarray}
    x &= C \sin (\Phi_D - \Phi_C) \, ,  \\
    y &= C \cos (\Phi_D - \Phi_C) \, , \\
    z &= D.
\end{eqnarray}
\endnumparts
Using the Jacobi--Anger expansion for all three terms in the integrand, we get a triple sum. Gathering the non-vanishing terms the integral solution can be written as:
\begin{eqnarray}
    I_{\alpha} = &e^{L_0} + I_0(x) I_0(y) I_0(z) + \nonumber\\
    & 2 I_0(x) \sum_{n=1}^\infty I_n(z) I_{2n}(y) + 2 I_0(z) \sum_{n=1}^\infty I_{2n}(y) I_{2n}(x) (-1)^n + \nonumber\\
    & 2 I_0(y) \sum_{n=1}^\infty I_{n}(z) I_{2n}(x) (-1)^n + 2 \sum_{n=1}^\infty I_{2n}(z) I_{2n}(y) I_{2n}(x) (-1)^n + \nonumber\\
    & 2 \sum_{n=1}^\infty I_{n}(z) I_{4n}(y) I_{2n}(x) (-1)^n + 2 \sum_{n=1}^\infty I_{n}(z) I_{2n}(y) I_{4n}(x),
\end{eqnarray}
\setcounter{footnote}{0}
where $I_n$ are modified Bessel functions of the first kind. The first non-trivial term comes from when we gather the constants from all three sums. The next three terms are from combinations when we pick the constant from one of the sums and $\cos (n\Phi_{\alpha})$ from the other two, which results in a non-vanishing integral of cosine squared. The last three terms are from when two $\cos (n\Phi_{\alpha})$ terms are combined with a $\cos (2n\Phi_{\alpha})$. While the sums go to infinity, these can be well approximated by just the first few terms for relatively low signal amplitudes. Using fast implementations of the Bessel functions\footnote{We use the implementation in \texttt{scipy} \cite{scipy} compiled in \texttt{Numba} \cite{numba_paper, numba_zenodo} via the \texttt{numba-scipy} package (\url{https://github.com/numba/numba-scipy})} makes the calculation of the phase marginalized likelihood cost less than two times more than a single non-marginalized likelihood (see Table \ref{tab:timings}). For large amplitudes, the series oscillates and convergence cannot be achieved due to numerical errors. However, in this regime, the integral can be approximated with the Laplace method \cite{high-order-laplace}:
\begin{eqnarray}
    I_{\alpha} \simeq \frac{\exp \left[ L_0+f_0 \right]}{\sqrt{2\pi |f_2|}} \Bigg( &1 + \frac{f_4}{8|f_2|^2} + \frac{5f_3^2}{24|f_2|^3} + \frac{f_6}{48|f_2|^3} +\nonumber\\
    &\frac{35f_4^2}{384|f_2|^4} + \frac{7f_3f_5}{48|f_2|^4} + \frac{35f_3^2f_4}{64|f_2|^5} + \frac{385f_3^4}{1152|f_2|^6} \Bigg),
    \label{eq:Laplace}
\end{eqnarray}
where $f_n$ is the $n$th derivative of $f(\Phi)$ w.r.t.~$\Phi$. The first term is the usual Gaussian integral, while terms in the parentheses account for non-vanishing higher order derivatives of the function. The maximum is found by searching for a root of $f_1$ on both the $(0, \pi)$ and $(\pi, 2\pi)$ intervals using Brent's method \cite{Brents_method} as implemented in \texttt{scipy.optimize.brentq} \cite{scipy}. Then the real maximum can be picked from the two by requiring that $f_2<0$.

Thus a combination of these two methods can provide an accurate and efficient integration of phase parameters over the entire prior range of signal amplitudes. We tested this method against a simple numerical integration and found that the error these methods introduce in the log likelihood is less than $10^{-3}$, thus smaller than the errors introduced by inner product interpolation discussed in Section \ref{sec:interpolation}. This is sufficient for current datasets, but if higher precision is required, the Laplace method in Eq.~(\ref{eq:Laplace}) can be continued to higher order.

Note also that while this phase integration is presented as part of this new approach that uses the interpolated inner products, the same marginalization could be introduced in any other method where the $N^i_{\alpha}$ and $M^{ij}_{\alpha}$ matrices are available, e.g.~in \texttt{QuickCW}.

\subsection{Numerical pulsar distance marginalization}
\label{sec:distance-marg}

After the semi-analytic phase marginalization, we can also marginalize over pulsar distances numerically. This is relatively cheap, because the pulsar distance only affects the signal in the given pulsar, and even for moderate frequency ($\sim$20 nHz) and high chirp mass ($\sim10^9\ M_{\odot}$) the likelihood is slowly changing and can be integrated out over just a few points (we default to 10). In addition, making 10 calls to the likelihood at different pulsar distances (and thus pulsar frequencies) does not increase the cost 10-fold, since only part of the calculation needs to be repeated. Similar to previous studies, we adopt a Gaussian prior on pulsar distances based on the measured value and its uncertainty. The following method can trivially be extended to more sophisticated prior shapes like the ones introduced in Ref.~\cite{nanograv_12p5yr_cw}. To make best use of our likelihood calls, we can rewrite the integral as a sum over points evenly spaced according to the CDF of the Gaussian prior:
\begin{eqnarray}
    p(d|\theta) &= \int_0^\infty p(d|\theta,L) p(L) \ dL = \int_0^1 p(d|\theta,L) \ dP(L)\nonumber\\ &\approx \frac{1}{N} \sum_{i=1}^N p\Bigg(d\Bigg|\theta,L=P^{-1}\left[\frac{2i-1}{2N}\right]\Bigg),
\end{eqnarray}
where $L$ is the distance to the pulsar, $p(L)$ is the PDF and $P(L)$ is the CDF of the distance prior. We can see from Table \ref{tab:timings} that doing this sum over 10 points makes the likelihood call about 7 times more expensive as the phase-marginalized likelihood. We argue that this is well-worth it, since now our likelihood costs 10 times the original interpolated likelihood (still $\sim$400 times faster than a regular \texttt{enterprise} likelihood call), but now we only need to sample over $8$ parameters, instead of $8+2N_{\rm PSR}$ parameters, which for recent PTA datasets with dozens of pulsars is a significant reduction in parameter dimension\footnote{E.g. for the NANOGrav 15yr dataset, this reduces the number of parameters from 142 to 8.}. Note that this is essentially the same idea as in Ref.~\cite{Steve_accelerated_CW}, except they draw values randomly from the prior of $L$ instead of using a uniform-in-probability grid via the CDF.

\section{Tests on simulated datasets}
\label{sec:test}
\begin{figure}[htb]
 \centering
   \includegraphics[width=1\textwidth]{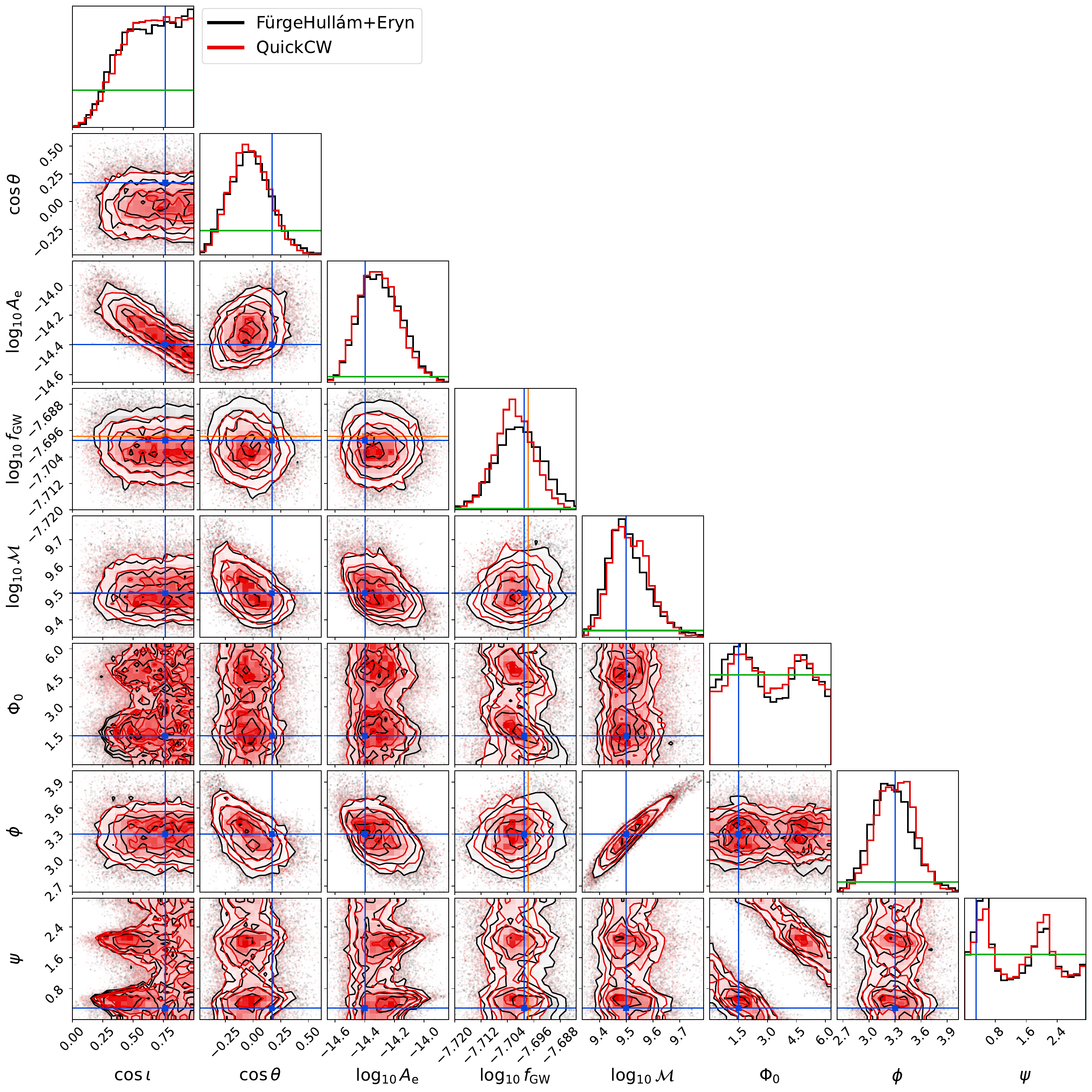}
   %\fakeimage
 \caption{Posterior comparison of analysis sampling the splined phase-and-distance-marginalized \texttt{FürgeHullám} likelihood with the \texttt{Eryn} ensemble sampler (6 min run, black) and a QuickCW analysis (50 hour run, red). Green lines indicate the prior distributions, and blue shows the injected model parameters. Orange shows the injected frequency if we reference that at the middle of the 16 year observing window instead of referencing it at the first TOA.}
 \label{fig:big_corner}
\end{figure}

To validate the parameter estimation accuracy of this new algorithm, we tested it on simulated datasets made to resemble the NANOGrav 15yr narrow-band dataset \cite{nanograv_15yr_data}. Our datasets have the same pulsars, same observing epochs and number of TOAs. The white noise and red noise in each pulsar are simulated with the best-fit parameters found for the real data \cite{nanograv_15yr_data, nanograv_15yr_noise}. We also added a simulated SMBHB with $f_{\rm E}=20$ nHz, $\log_{10}\mathcal{M}=9.5$, $A_{\rm e}=4\times10^{-15}$, which resulted in a signal-to-noise ratio of SNR=11.4. Figure \ref{fig:big_corner} shows the 1 and 2-dimensional posterior distributions via the \texttt{FürgeHullám} likelihood sampled with the \texttt{Eryn} \cite{eryn} ensemble-sampler (black), and a standard \texttt{QuickCW} run for comparison (red). Green lines indicate the prior distributions, and blue shows the injected model parameters. We used uniform priors on all eight parameters with the following boundaries: $\cos \iota \in [-1,1]$, $\cos \theta \in [-1,1]$, $\log_{10} A_{\rm e} \in [-18,-11]$, $\log_{10} f_{\rm GW} \in [-8.704,-7.5]$, $\log_{10} \mathcal{M} \in [7,10]$, $\Phi_0 \in [0, 2\pi]$, $\phi \in [0,2\pi]$, $\Psi \in [0,\pi]$, where the lower bound on $\log_{10} f_{\rm GW}$ corresponds to $f_{\rm GW}=1/T_{\rm obs}$. \texttt{QuickCW} was run for 1 billion fast steps, 100 thousand slow steps, which took about 50 hours. \texttt{Eryn} was run with 100 walkers for 200 iterations after a 100 iteration burnin (initialized around the injected location), and took about 6 minutes to run (after the 4.5 hour setup discussed above). While QuickCW produces many more samples, due to the larger autocorrelation length, the two analyses resulted in roughly the same number of independent samples.

We can see on Figure \ref{fig:big_corner} that even in this large-SNR highly-evolving signal limit, the two results are nearly identical. The most significant (but still minor) difference can be seen in the recovery of the GW frequency parameter. This is due to the fact that \texttt{FürgeHullám} ignores the frequency evolution of the signal within the observing timespan. Since the frequency is referenced at the first TOA, and the signal chirps up in frequency, \texttt{FürgeHullám} recovers a wider posterior centered slightly higher. We also show the injected frequency if we reference that at the middle of the 16 year observing window, the offset of which is consistent with the shift in posterior.

Ignoring the signal evolution within the observing window is expected to have the strongest effect for such a high-frequency high-chirp-mass and high-SNR signal as this one. To quantify this, we can compare the systematic bias we get by assuming no evolution with the statistical error in the measured frequency. The bias can be associated with the amount of frequency evolution during half the observing window:
\begin{equation}
\Delta f_{\rm bias} \approx \frac{96}{5} \mathcal{M}^{5/3} (\pi f_{\rm E})^{8/3} f_{\rm E} T_{\rm obs}/2,
\end{equation}
where we only kept the first-order linear term in the frequency evolution. The statistical error in $f_{\rm E}$ is observed to be 0.27 nHz in the example shown on Figure \ref{fig:big_corner}, and we know that it follows a simple scaling:
\begin{equation}
\Delta f_{\rm stat} \propto \frac{1}{{\rm SNR} T_{\rm obs}}.
\end{equation}
Thus the ratio of these two can be expressed as:
\begin{equation}
\frac{\Delta f_{\rm bias}}{\Delta f_{\rm stat}} \approx 0.011 \left(\frac{\mathcal{M}}{10^9 \ M_{\odot}}\right)^{5/3} \left(\frac{f_{\rm E}}{20 \ {\rm nHz}}\right)^{11/3}  \left( \frac{T_{\rm obs}}{10 \ {\rm yr}} \right)^2 \left( \frac{\rm SNR}{10} \right).
\end{equation}
A reasonable requirement is that the bias should be smaller (preferably much smaller) than the statistical error. Substituting SNR=11.4 and $T_{\rm obs}=16$ yr, this gives a ratio of 0.22, which is consistent with the slight bias in recovered frequency we have seen on Figure \ref{fig:big_corner}. Binaries are expected to be most numerous at lower frequencies (see e.g.~\cite{RosadoExpectedProperties, Luke_single_source, RealisticDetection1}), and we do not expect SNRs much higher than 10 in the near future. Thus we can see that the ratio of bias and statistical error stays low for most realistic scenarios, and thus ignoring evolution within the observing window is well justified. However, one needs to be careful at high observer-frame chirp mass and frequency values and as the observing time and SNR increases.

We also tested the model selection capabilities of this new approach. To do so we analyzed the same dataset as above, except with a lower amplitude to achieve a more realistic SNR of 6.7. We are interested in the Bayes-factor between a model including an individual binary and pulsar noise (white and red noise), and a model including only pulsar noise. We calculate this Bayes factor using two different methods: i) with the \texttt{FürgeHullám} likelihood using nested sampling (NS, \cite{nested_sampling}) via \texttt{dynesty}\cite{dynesty}; ii) using a \texttt{QuickCW} run via the Savage-Dickey density ratio (SD, \cite{SD_BF}). The former took 30 mins and resulted in the estimate of $\ln {\rm BF}_{NS}=5.84\pm0.24$, while the latter took 50 hours to run and resulted in the estimate of $\ln {\rm BF}_{SD}=5.82\pm0.11$. So we can see that the new likelihood with NS gives a consistent and comparably accurate Bayes factor as the frequently used SD method in a fraction of the time.
%A=7e-16

\section{Conclusion and future work}
\label{sec:conclusion}

In this paper we presented a new approach to analyzing signals from individual SMBHBs in PTA data, which provides superior speed compared to current analysis techniques as long as the noise models are held fixed. Thus in its current implementation this approach can be viewed as either an efficient inference and model selection tool or as a first-pass quick search, which complements efficient full noise-marginalized search algorithms like \texttt{QuickCW}. It will also help us keep up with growing datasets, which require more and more efficient algorithms to keep the analysis tractable. Increased efficiency can also be beneficial in that we have no incentive to exclude existing data due to computational limitations, as long as the noise can be sufficiently understood.

We demonstrated that after an initial setup this new approach can provide efficient parameter estimation and model selection 100-1000 times faster than current methods. The speedup is still a factor of 10-100 when taking into account the setup time. In addition, the same setup can be used for the analysis of any deterministic signal that can be expressed as a sum of sines and cosines. These will be explored in more details in future studies, but some possibilities include:
\begin{itemize}
    \item Comparison with alternative sinusoidal models, like incoherent sine model, monopolar or dipolar sine model;
    \item Non-GW sinusoidal models like ultralight dark matter (see Ref.~\cite{nanograv_15yr_new_physics} and references therein);
    \item Repeated runs on sky scrambled/phase shifted datasets for false-alarm estimation similar to those used for the GWB (see e.g.~Refs.~\cite{neil_robust_detection,all_correlations_must_die});
    \item Eccentric binary analysis \cite{steve_eccbin,abhimanyu_eccbin_2020,abhimanyu_eccbin_2022,nanograv_12p5yr_eccbin};
    \item Alternative polarization modes \cite{logan_alt_pol_cw};
    \item Multiple binaries \cite{babak_sesana_multiple_cw,BayesHopper}.
\end{itemize}
In addition, this new approach will particularly beneficial for large simulation studies, since only a small fraction of the setup needs to be repeated if only the data changes between realizations, but the observing epochs and covariance matrix stays the same.

While the current implementation only works if the noise is held fixed due to the inner product interpolation, the other two components of this approach could benefit search algorithms like \texttt{QuickCW} as well. In addition, there are potential ways to remedy the fixed-noise limitation of this pipeline to make it a full search algorithm in the future. One possibility would be to change the $M^{ij}_{\alpha}$ interpolation from using a 2-dimensional grid over all frequencies to a 1-dimensional one covering possible pulsar frequencies and recalculate them each time we change the Earth-term frequency. That would still require 100s of inner product calculations for each update, but potentially with GPU-based parallelization it can be done on the fly, which would also allow for changing the noise parameters. This can also be thought of as an extension of the \texttt{QuickCW} approach, where all parameters except the Earth-term frequency become fast projection parameters. This approach will be explored in more detail in a future study.

Another way to marginalize over parameters of the covariance matrix (in particular a common red noise process) would be to directly sample the Fourier coefficients instead of using the marginalized covariance matrices. This has been proposed before \cite{lentati_2013}, but have not been used widely because the computation via the marginalized covariance matrix was more efficient. However, there have been recent developments in efficient sampling of these coefficients \cite{NimaGibbs,RutgerCoefficientSampling}, which could change this situation. In particular, the fact that we precalculate sine cosine inner product for the deterministic model anyway, would further help in combining these efforts. In addition, it has also been suggested, that this is the way forward for deterministic signal searches. This is because directly sampling the GWB Fourier coefficients allows us to keep the likelihood factorized (see Eq.~(\ref{eq:log_like})) even under a correlated background model (see details of the idea in Appendix A in Ref.~\cite{QuickCW} and a first implementation in Ref.~\cite{AidenPaper}), which is a requirement for both \texttt{QuickCW} and the \texttt{FürgeHullám} approach presented in this paper.

The assumption of no inter-pulsar correlations could also be relaxed by introducing additional terms in the log likelihood in Eq.~(\ref{eq:log_like}). These would correspond to cross-terms between data and/or signal in different pulsars. Such an approach would mean precalculating and storing about a factor of $N_{\rm PSR}$ more inner products. In addition, some terms would depend on phases in different pulsars, which would introduce 2-dimensional integrals when marginalizing over the phase. The benefit of this idea is that we can avoid sampling the GWB coefficients. Whether this approach is feasible and efficient enough to be competitive with the Fourier coefficient sampling method will be examined in future work.

\ack
The author thanks Neil Cornish and members of the OSU Gravity Group and the NANOGrav Collaboration for fruitful discussions throughout this project. We would also like to thank Bjorn Larsen and Sarah Burke Spolaor for feedback on the manuscript. The author is also grateful to an anonymous contributor to Mathematics Stackexchange for the idea that helped identify the Laplace method as an appropriate approximation of the pulsar phase integral in the large-amplitude limit, see Eq.~(\ref{eq:Laplace}).
We appreciate the support of the NSF Physics Frontiers Center Award PFC-2020265.

\section*{References}
\bibliography{cwcoherence}{}
\bibliographystyle{unsrt_et_al}

\end{document}